\documentstyle[preprint,eqsecnum,aps,epsbox]{revtex}
\tighten

\begin{document}

\draft


\preprint{YITP-99-61, gr-qc/9910013}

\title{Is the brick-wall model unstable for a rotating background?}
\author{Shinji Mukohyama}
\address{
Yukawa Institute for Theoretical Physics, 
Kyoto University \\
Kyoto 606-8502, Japan \\
Department of Physics and Astronomy, 
University of Victoria\\ 
Victoria, BC, Canada V8W 3P6
}
\date{\today}

\maketitle


\begin{abstract} 

The stability of the brick wall model is analyzed in a rotating
background. It is shown that in the Kerr background without horizon
but with an inner boundary a scalar field has complex-frequency modes
and that, however, the imaginary part of the complex frequency can be
small enough compared with the Hawking temperature if the inner
boundary is sufficiently close to the horizon, say at a proper
altitude of Planck scale. Hence, the time scale of the instability due
to the complex frequencies is much longer than the relaxation time
scale of the thermal state with the Hawking temperature. Since ambient 
fields should settle in the thermal state in the latter time scale,
the instability is not so catastrophic. Thus, the brick wall model is
well defined even in a rotating background if the inner boundary is
sufficiently close to the horizon. 

\end{abstract}



\section{Introduction}

Understanding the origin of black hole entropy is one of the most
interesting problems in black hole physics. The black hole entropy is 
given by the Bekenstein-Hawking formula~\cite{Bekenstein,Hawking} as 
%
\begin{equation}
 S_{BH} = \frac{1}{4}A_H,
\end{equation}
where $A_H$ is area of the horizon.

It seems that full understanding of black hole entropy requires the
theory of quantum gravity, which we do not know yet. However, we
believe that general feature of black hole entropy can be understood
by semiclassical theory, namely, quantum field theory in a fixed
gravitational background. In fact, the brick wall model proposed by
'tHooft~\cite{tHooft} succeeded to derive the proportionality of black 
hole entropy to the horizon area by identifying the black hole entropy 
with thermal entropy of ambient quantum fields raised to the
Hawking temperature. It was recently clarified that in this model
backreaction is small enough and that this model is actually a
self-consistent model as a semiclassical
theory~\cite{Mukohyama&Israel}. Moreover, it was shown that this model
seeks the maximal value of entanglement entropy in the space of states
whose backreaction is small enough~\cite{Mukohyama}.

Originally, the brick wall model is proposed in the spherically
symmetric, static background, say, the Schwarzschild
background. Hence, it seems interesting to see how this model is
extended to a rotating background, say, the Kerr 
background~\cite{Lee&Kim,HKPS,Ho&Kang,Frolov&Fursaev}. 
However, it is known that in a rapidly rotating spacetime without
horizon a field has complex-frequency
modes~\cite{Sato&Maeda,Comins&Schutz} and that there is the so called
ergoregion instability~\cite{Kang}. Thus, it might be expected that
the brick-wall model in rotating background might be unstable and
unsuitable for the origin of black hole entropy.

In this paper we analyze the stability of the brick wall model in a
rotating background. We show that the time scale of the ergoregion
instability is much longer than the relaxation time scale of the
thermal state with the Hawking temperature. In the latter time scale
ambient fields should settle in the thermal state. Thus, the brick
wall model is well defined even in a rotating background.

In Sec.~\ref{sec:rotating-background} we summarize a quantum field
theory of a real scalar field in a $n$-dimensional axisymmetric
stationary spacetime to show how the appearance of complex-frequency
modes alters the structure of the quantum field theory. In
Sec.~\ref{sec:complex-frequency} we consider a scalar field in the
$4$-dimensional Kerr spacetime without horizon but with an inner
boundary to show the existence and a property of the complex
frequency. Section ~\ref{sec:summary} is devoted to summarize this
paper.


\section{Scalar field in rotating background}
	\label{sec:rotating-background}

Let us consider a general $n$-dimensional axisymmetric stationary
spacetime ${\cal M}$, whose metric is given by 
%
\begin{equation}
 ds^2 = -N^2dt^2 + \rho^2(d\varphi - \omega_B dt)^2 
	+ q_{ab}dx^adx^b, \label{eqn:metric}
\end{equation}
where $a,b=1,2,\cdots,(n-2)$. 
Here, the lapse function $N$, the Bardeen angular velocity (or minus
the $\varphi$-component of the shift vector) $\omega_B$,
$(\varphi\varphi)$-component $\rho^2$ of the metric and the 
$(n-2)$-dimensional metric $q_{ab}$ are assumed to depend only on the
$(n-2)$-dimensional coordinates $\{x^a\}$. 
On this background spacetime we consider a real scalar field $\phi$
described by the action
%
\begin{equation}
 I = -\frac{1}{2}\int_{\cal M}d^nx\sqrt{-g}
	(g^{\mu\nu}\partial_{\mu}\phi\partial_{\nu}\phi +
	\mu^2\phi^2),
	\label{eqn:action}
\end{equation}
where the mass $\mu$ of the field can depend only on the
$(n-2)$-dimensional coordinates $\{x^a\}$.
We impose the boundary condition 
%
\begin{equation}
 N\rho\phi n^{\mu}\partial_{\mu}\phi = 0 
	\quad \mbox{on} \quad \partial{\cal M},
\end{equation}
where the boundary $\partial{\cal M}$ of the spacetime ${\cal M}$ is
supposed to be invariant under translations generated by Killing
vectors $\partial_t$ and $\partial_{\varphi}$, and $n^{\mu}$ is a unit 
normal to $\partial{\cal M}$. 
Note that a part of $\partial{\cal M}$ can be taken at spatial
infinity.

In this paper we quantize the system of the scalar field with respect
to the time evolution vector $D$ defined by  
%
\begin{equation}
 D = \left(\frac{\partial}{\partial t}\right)_{\varphi,x^a}
	+ \Omega (x^a)
	\left(\frac{\partial}{\partial\varphi}\right)_{t,x^a}, 
\end{equation}
where $\Omega(x^a)$ is an arbitrary function of $\{x^a\}$ such that
$D$ is timelike in ${\cal M}$. 
For this choice of the time evolution vector, it is convenient to use
a new coordinate system $(t,\tilde{\varphi},x^a)$ defined by
%
\begin{equation}
 \tilde{\varphi} = \varphi - \Omega(x^a) t
\end{equation}
since for this coordinate system
%
\begin{equation}
 D = \left(\frac{\partial}{\partial t}\right)_{\tilde{\varphi},x^a}.
\end{equation}
Following the usual quantization procedure, it turns out that the
canonical momentum $\pi$ conjugate to $\phi$ and, thus, the equal-time 
commutation relations are independent of the choice of $\Omega(x^a)$:
%
\begin{equation}
 \pi \equiv \frac{\delta L}{\delta(D\phi)} = 
	\frac{\rho\sqrt{q}}{N}\left[
	D\phi + (\omega_B - \Omega )\left(
	\frac{\partial\phi}{\partial\tilde{\varphi}}\right)_{t,x^a}
	\right]
	= \frac{\rho\sqrt{q}}{N}\left[
	\left(\frac{\partial\phi}{\partial t}\right)_{\varphi,x^a} +
	\omega_B\left(\frac{\partial\phi}{\partial\varphi}\right)_{t,x^a}
	\right],
\end{equation}
where the Lagrangian $L$ is defined by $I=\int dtL$. Therefore, the
quantization procedure we follow is independent of the choice of the
time evolution vector $D$. In this sence there is no ambiguity in the
quantization.

Off course, there are much freedom in selecting a ground state: we
have freedom in the choice of a set of positive-frequency mode
functions. In the following, we give one example of the choice of the
set of positive-frequency mode functions by using a separation of
variables. Other choices give different ground states. However, the
Hilbert space of all quantum state is independent of the choice of the 
set. For example, one ground state can be expressed as excited states
above other ground states.

To quantize the system of the scalar field we raise the field $\phi$
to an operator and decompose it by mode functions:
%
\begin{equation}
 \phi = \sum_{(\omega lm)\in P}
	\left(\Phi_{\omega lm}a_{\omega lm} +
	\Phi_{\omega lm}^*a_{\omega lm}^{\dagger}\right),
	\label{eqn:phi-decomp}
\end{equation}
where the set $P$ and the mode functions $\{\Phi_{\omega lm}\}$ will 
be defined below by Eq.~(\ref{eqn:def-P}) and Eq.~(\ref{eqn:def-Phi}),
respectively.

In order to define the mode functions $\{\Phi_{\omega lm}\}$ in the
above expansion, let us seek solutions $\{\Psi_{\omega lm}\}$ of the
field equation by the following separation of variables. 
%
\begin{equation}
 \Psi_{\omega lm} = f_{\omega lm}(x^a)e^{-i\omega t}e^{im\varphi}.
	\label{eqn:separation}
\end{equation}
The function $f_{\omega lm}(x^a)$ is a solutions of the equation
%
\begin{equation}
 \frac{1}{N\rho\sqrt{q}}\partial_a(N\rho\sqrt{q}q^{ab}
	\partial_b f_{\omega lm}) 
	+\left[\frac{(\omega -\omega_B m)^2}{N^2} 
	-\frac{m^2}{\rho^2} - \mu^2\right] f_{\omega lm} =0, 
	\label{eqn:EOM-S}
\end{equation}
with the boundary condition
%
\begin{equation}
 N\rho f_{\omega lm} n^{\mu}\partial_{\mu} f_{\omega' lm} = 0
	\quad\mbox{on}\quad \partial{\cal M},
\end{equation}
where $n^{\mu}$ denotes a unit normal to the boundary 
$\partial{\cal M}$.  
(A part of $\partial{\cal M}$ can be taken at spatial infinity.)
Here note that, because of the invariance of Eq.~(\ref{eqn:EOM-S})
under $(\omega,m)\leftrightarrow (-\omega,-m)$, we can assume that 
%
\begin{equation}
 f^*_{\omega lm} = f_{-\omega^*,l,-m}.\label{eqn:symmetry1}
\end{equation}
We can choose the quantum number $l$ so that 
%
\begin{equation}
 \int d^{n-2}x\frac{\rho\sqrt{q}}{N}(\omega-\omega_Bm)
	f_{\omega lm}f^*_{\omega^* l'm} = 0 
	\quad\mbox{unless}\quad l=l'. \label{eqn:normalization1}
\end{equation}
With this choice of the quantum number $l$, the following property
holds. 
%
\begin{eqnarray}
 (\Psi_{\omega lm},\Psi_{\omega' l'm'})_{KG} = 0 
	\quad\mbox{unless}\quad 
	\omega^*=\omega',l=l'\ \mbox{and}\ m=m',
\end{eqnarray}
where the Klein-Gordon norm $(\Phi,\Psi)_{KG}$ is given by
%
\begin{equation}
 (\Phi,\Psi)_{KG} = -i\int d^{n-1}x\frac{\rho\sqrt{q}}{N}
	(\Phi D\Psi^* - \Psi^* D\Phi). 	\label{eqn:KG-norm}
\end{equation}

Now the set $P$, over which the summation is taken in
Eq.~(\ref{eqn:phi-decomp}), is defined as 
%
\begin{eqnarray}
 P & = & P_R \cup P_C,\label{eqn:def-P}\\
 P_R & = & \{(\omega lm)| \omega \ \mbox{is real},
	(\Psi_{\omega lm},\Psi_{\omega lm})_{KG}>0\},\nonumber\\
 P_C & = & \{(\omega lm)| \Im\omega > 0\}\nonumber.
\end{eqnarray}
The mode functions $\{\Phi_{\omega lm}\}$ in the expansion
(\ref{eqn:phi-decomp}) are defined by 
%
\begin{eqnarray}
 \Phi_{\omega lm} & = & 
	\frac{1}{\sqrt{C_{\omega lm}}}\Psi_{\omega lm}
	\quad\mbox{for}\quad (\omega lm)\in P_R,\nonumber\\
 \Phi_{\omega lm} & = & 
	\frac{1}{\sqrt{2C_{\omega lm}}}
	(\Psi_{\omega lm}+e^{i\alpha_{\omega lm}}\Psi_{\omega^*lm})
	\quad\mbox{for}\quad (\omega lm)\in P_C,\label{eqn:def-Phi}
\end{eqnarray}
where the real constants $C_{\omega lm}$ ($>0$) and $\alpha_{\omega
lm}$ are defined by 
%
\begin{equation}
 (\Psi_{\omega lm},\Psi_{\omega^* lm})_{KG} = 
	C_{\omega lm}e^{i\alpha_{\omega lm}}.
\end{equation}
For the above definition of $P$ and $\{\Phi_{\omega lm}\}$, the
following property can be easily derived.
%
\begin{eqnarray}
 (\Phi_{\omega lm},\Phi_{\omega'l'm'})_{KG} & = & 
	\delta_{\omega\omega'}\delta_{ll'}\delta_{mm'},
	\label{eqn:Phi-Phi=delta}\\
 (\Phi_{\omega lm},\Phi^*_{\omega'l'm'})_{KG} & = & 0
	\label{eqn:Phi-Phi*=0}\\
 (\Phi^*_{\omega lm},\Phi^*_{\omega'l'm'})_{KG} & = & 
	-\delta_{\omega\omega'}\delta_{ll'}\delta_{mm'},
	\label{eqn:Phi*-Phi*=-delta}
\end{eqnarray}
for $\forall (\omega lm)\in P$ and $\forall (\omega'l'm')\in P$. It is 
these properties that lead us to the above definition of $P$ and 
$\{\Phi_{\omega lm}\}$. In Appendix~\ref{app:KGnorm}, it is shown that 
the local integrability of Eq.~(\ref{eqn:EOG}) below requires
Eq.~(\ref{eqn:Phi-Phi*=0}) and that the normalizability of the ground
state $|0\rangle$ requires the left hand side of
Eq.~(\ref{eqn:Phi-Phi=delta}) to be positive definite as a matrix with 
arguments $\lambda=(\omega lm)$ and $\lambda'=(\omega'l'm')$. 
(Eq.~(\ref{eqn:Phi*-Phi*=-delta}) is an immediate consequence of
Eq.~(\ref{eqn:Phi-Phi=delta}).)

Since the above orthonormality of the mode functions imply 
%
\begin{equation}
 [a_{\lambda},a^{\dagger}_{\lambda'}]=\delta_{\lambda\lambda'},\quad 
 [a_{\lambda},a_{\lambda'}]=0,
\end{equation}
the Hilbert space ${\cal F}$ of all quantum states can be constructed
as a symmetric Fock space spanned by the states 
$|\{N_{\lambda}\}\rangle$ defined by  
%
\begin{equation}
 |\{N_{\lambda}\}\rangle = 
	\prod_{\lambda\in P}
	\frac{(a_{\lambda}^{\dagger})^{N_{\lambda}}}
	{\sqrt{N_{\lambda}!}}	|0\rangle, 
	\label{eqn:Fock-space}
\end{equation}
where the ground state $|0\rangle$ is defined by 
%
\begin{equation}
 a_{\lambda}|0\rangle = 0 \quad \mbox{for}\quad 
	\forall \lambda\in P.	\label{eqn:EOG}
\end{equation}
Hereafter, $\lambda$ denotes $(\omega lm)$ and $\bar{\lambda}$ denotes 
$(-\omega^*,l,-m)$.

The canonical Hamiltonian $H[D]$ with respect to the time evolution
vector $D$ is given by 
%
\begin{equation}
 H[D] = \frac{i}{2}(D\phi,\phi)_{KG}. 
\end{equation}
Hence, if $\Omega$ is a constant~\footnote{
If $\Omega$ is not a constant then $H[D]$ is not a conserved quantity 
in general since $\Omega\phi$ does not satisfy the equation of
motion.
}, 
then $H[D]$ is a conserved quantity and can be expressed as  
%
\begin{equation}
 H[D] = \frac{1}{2}\sum_{\lambda\in P}\left[
	(\Re\omega -\Omega m)(a_{\lambda}a^{\dagger}_{\lambda}
	+ a^{\dagger}_{\lambda}a_{\lambda})
	+ i\Im\omega
	(a^{\dagger}_{\lambda}a^{\dagger}_{\bar{\lambda}}
	- a_{\lambda}a_{\bar{\lambda}})\right].
	\label{eqn:H[D]}
\end{equation}
Note that any states of the form (\ref{eqn:Fock-space}) are not
eigenstates of this Hamiltonian unless all $\omega$ are real. 
(Off course, if there is no complex $\omega$ then all states of the
form (\ref{eqn:Fock-space}) are eigenstates of this Hamiltonian.) 
Moreover, in Appendix~\ref{app:complex-frequency-mode} it is shown
that there is no ground state suitable for this Hamiltonian unless all
$\omega$ are real. To be precise, it is always impossible to eliminate 
terms including $a_{\lambda}^{\dagger}a_{\bar{\lambda}}^{\dagger}$
or $a_{\lambda}a_{\bar{\lambda}}$ in (\ref{eqn:H[D]}) by a Bogoliubov
transformation. (See Appendix~\ref{app:complex-frequency-mode}.)
Thus, if there is a complex $\omega$ then there is no stable ground
state in ${\cal F}$. Hence, existence of complex-frequency modes imply 
a kind of instability in quantum field theory. This conclusion is
consistent with the results of Refs.~\cite{Schroer&Swieca,Kang} that
spectrum of the Hamiltonian becomes continuous and that eigen states
are not normalizable if there is a complex-frequency mode.

Off course, there is a corresponding instability in classical
theory: if there is a complex frequency with positive imaginary part
then a solution expressed as (\ref{eqn:separation}) grows
exponentially in time. On the other hand, if imaginary part of
frequency is negative then the corresponding solution decays in time
but grows exponentially in inverse time.

Therefore, in both classical and quantum senses, appearance of
complex-frequency modes implies instability of the system. In the
next section, we show that in the brick wall model a scalar field has
complex-frequency modes. Hence, it might be expected that the brick
wall model might be unstable and that it might be an unsuitable model
to seek entropy for an equilibrium state. However, it turns out that
the imaginary part of the complex frequency can be made arbitrarily
small by making the inner boundary close enough to the horizon. In
fact, in the next section, we show that the imaginary part is small
enough compared with the Hawking temperature if the inner boundary is
sufficiently close to the horizon, say at a proper altitude of Planck
scale.


\section{Complex frequency modes and stability of the brick wall
model} 
	\label{sec:complex-frequency}

For simplicity, let us consider the ($4$ dimensional) Kerr spacetime
as a background and suppose that the mass $\mu$ is a non-zero
constant. The metric is given by 
%
\begin{equation}
 ds^2 = -\left(1-\frac{2Mr}{\Sigma}\right)dt^2 
	+ \frac{\Sigma}{\Delta}dr^2 + \Sigma d\theta^2
	+ R^2\sin^2\theta d\varphi^2
	- \frac{4Mar}{\Sigma}\sin^2\theta d\varphi dt,
\end{equation}
where
%
\begin{eqnarray}
 \Sigma & = & r^2 + a^2\cos^2\theta,\nonumber\\
 \Delta & = & r^2 + a^2 - 2Mr,\nonumber\\
 \Sigma R^2 & = & (r^2 + a^2)^2 - a^2\Delta\sin^2\theta.
\end{eqnarray}
This is of the form (\ref{eqn:metric}) with 
%
\begin{eqnarray}
 N^2 & = & \frac{\Delta}{R^2},\nonumber\\
 \omega_B & = & \frac{2Mar}{\Sigma R^2},\nonumber\\
 \rho^2 & = & R^2\sin^2\theta,\nonumber\\
 q^{ab}dx^adx^b & = & \frac{\Sigma}{\Delta}dr^2+\Sigma d\theta^2. 
\end{eqnarray}

We only consider the region $r\geq r_0$ in this spacetime: we impose
the Dirichlet boundary condition on the field $\phi$ at
$r=r_0$. Hereafter we assume that $r_0>M+\sqrt{M^2-a^2}$: there is no
horizon in the region $r\geq r_0$.

\begin{figure}
 \begin{center}
  \epsfile{file=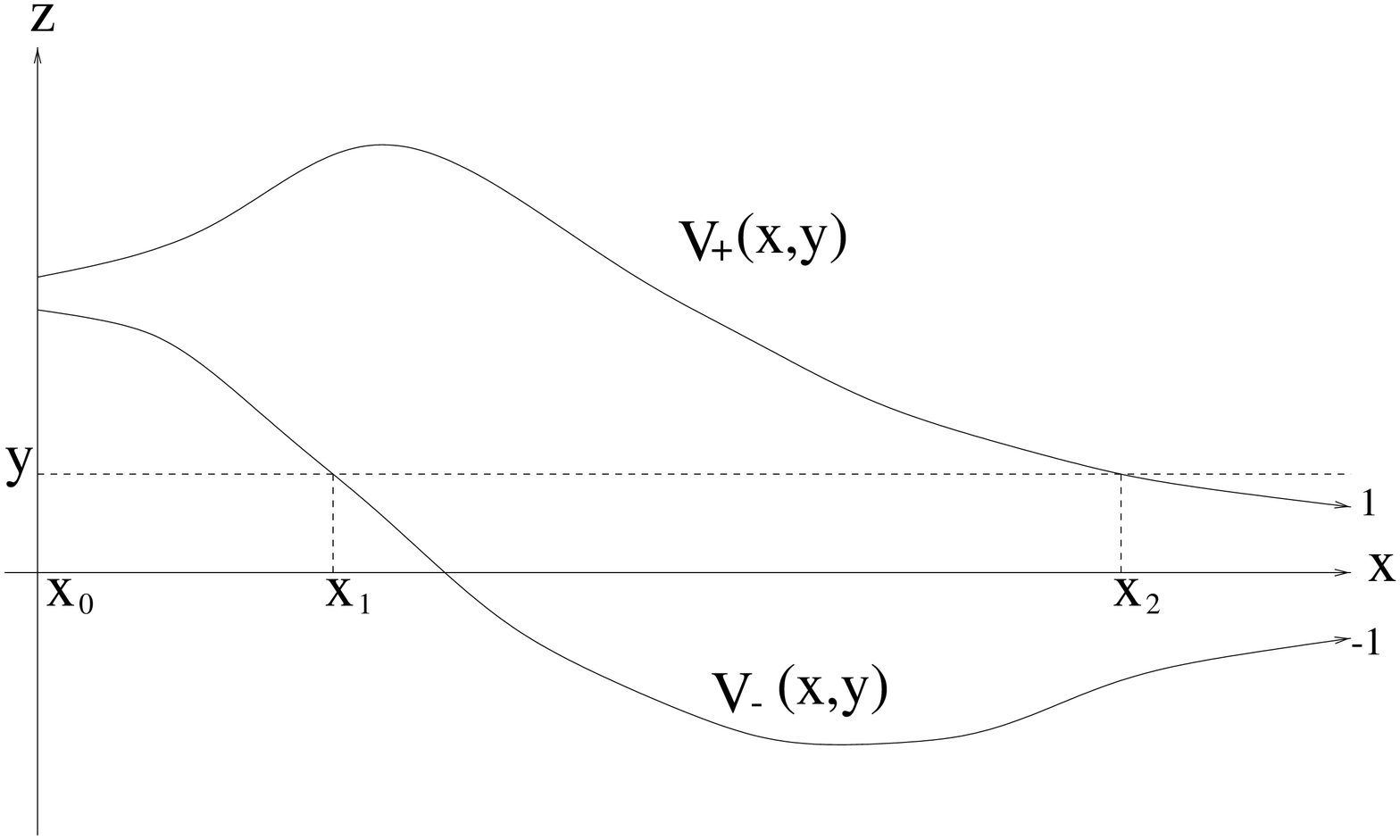,scale=0.5}
 \end{center}
\caption{
The typical form of the graphs $z=V_{\pm}(x,y)$ is written on a fixed
$y$ plane. Note that $V_{\pm}$ depend on $y$ as well as $x$ through
the eigen value $\lambda$ of equation (\ref{eqn:eq-S}). 
However, asymptotic behavior of $V_{\pm}$ in the limit $x\to\pm\infty$
does not depend on $y$. 
}
\label{fig:potential}
\end{figure}

It is well known that in this background Eq.~(\ref{eqn:EOM-S}) becomes
separable. In fact, we can find a solution of (\ref{eqn:EOM-S}) of the 
form
%
\begin{equation}
 f = \frac{u(r)}{\sqrt{r^2+a^2}}S(\theta),
\end{equation}
where $S$ satisfies
%
\begin{equation}
 \frac{1}{\sin\theta}\frac{d}{d\theta}
	\left(\sin\theta\frac{dS}{d\theta}\right)
	+ \left[\lambda - a^2\mu^2(y^2 - 1)\sin^2\theta
		-\frac{m^2}{\sin^2\theta}\right]S = 0,
	\label{eqn:eq-S}
\end{equation}
and the equation for $u$ can be written as 
%
\begin{equation}
 \frac{d^2u}{dx^2} + (y - V_+)(y - V_-) u =0
\end{equation}
by introducing the non-dimensional tortoise
coordinate $x$ by  
%
\begin{equation}
 \mu^{-1}\frac{dx}{dr} = \frac{r^2+a^2}{\Delta}, 
\end{equation}
or
%
\begin{equation}
 \mu^{-1}x = r + \frac{M}{\sqrt{M^2-a^2}}\left[
	r_+ \ln\left(\frac{r-r_+}{r_+-r_-}\right)
	- r_- \ln\left(\frac{r-r_-}{r_+-r_-}\right)\right]. 
\end{equation}
Here $r_{\pm}=M\pm\sqrt{M^2-a^2}$, $y\equiv\omega/\mu$ and $V_{\pm}$
are defined by 
%
\begin{eqnarray}
 \mu V_{\pm} & = &\frac{2maMr}{(r^2+a^2)^2} \pm
	\sqrt{\frac{\Delta\tilde{\mu}^2}{r^2+a^2}},\nonumber\\
 \tilde{\mu}^2 & = & \mu^2+\frac{\lambda}{(r^2+a^2)}
	-\frac{m^2a^2(r^2+a^2+2Mr)}{(r^2+a^2)^3}
	+\frac{2Mr+a^2}{(r^2+a^2)^2}
	-\frac{6Ma^2r}{(r^2+a^2)^3}. 
\end{eqnarray}
Note that in the horizon limit $r\to r_+$ (or $x\to -\infty$) both of
$V_{\pm}$ approach to the same value
%
\begin{equation}
 V_{\pm} \to \frac{ma}{2\mu Mr_+}, \label{eqn:limit-V}
\end{equation}
and that $V_{\pm}$ approach to $\pm 1$, respectively, in the limit
$r\to\infty$ (or $x\to\infty$). (See Fig.~\ref{fig:potential}.) 

Now let us seek complex frequency modes by examining a scattering
amplitude for real-frequency waves. The method we shall use here is
based on the following expected form of the scattering amplitude $S$
near a pole $y=y_R+iy_I$, providing that 
$|y_R|\gg |y_I|$~\cite{Comins&Schutz}. 
%
\begin{equation}
 S = e^{2i\delta_0}\times\frac{y-y_R+iy_I}{y-y_R-iy_I},
\end{equation}
where $\delta_0$ is a constant phase. Hence, if we can obtain this
form of a scattering amplitude by analyzing real-frequency waves then
we find an outgoing normal mode corresponding to $y=y_R+iy_I$ and an
incoming normal mode corresponding to $y=y_R-iy_I$. After that we
should confirm whether these normal modes converge or diverge in the
limit of $x\to\infty$. If these normal modes converge then they give
complex-frequency mode functions.

We first consider the case in which $ma>2\mu Mr_+$ and examine the
following five regime separately: (i) $1<y<V_-(x_0)$; (ii)
$y>V_+(x_0)$; (iii) $y<-1$; (iv) $V_-(x_0)\leq y\leq V_+(x_0)$; (v)
$-1\leq y\leq 1$.

(i) $1<y<V_-(x_0)$\\
In this regime, let the solution of $y=V_-$ and that of $y=V_+$ be 
$x=x_1(y)$ and $x=x_2(y)$, respectively. 
(See Fig.~\ref{fig:potential}.) 
In each region separated by $x_1$ and $x_2$, the WKB solution is given 
by 
%
\begin{eqnarray}
 u & = & \frac{C_1}{|T|^{1/4}}
	\sin\left(\int_{x_0}^x\sqrt{|T|}dx\right)
	\quad\mbox{for}\ x_0<x<x_1,	\nonumber\\
 u & = & \frac{C_2}{|T|^{1/4}}
	\exp\left(\int_{x_1}^x\sqrt{|T|}dx\right)
	+\frac{C_3}{|T|^{1/4}}
	\exp\left(-\int_{x_1}^x\sqrt{|T|}dx\right)
	\quad\mbox{for}\ x_1<x<x_2,	\nonumber\\
 u & = & \frac{C_4}{|T|^{1/4}}
	\exp\left(i\int_{x_2}^x\sqrt{|T|}dx\right)
	+\frac{C_5}{|T|^{1/4}}
	\exp\left(-i\int_{x_2}^x\sqrt{|T|}dx\right)
	\quad\mbox{for}\ x_2<x,
	\label{eqn:WKBsol1}
\end{eqnarray}
where
%
\begin{equation}
 T = (y-V_+)(y-V_-).
\end{equation}
Hence, the standard connection formula for WKB solutions gives
%
\begin{equation}
 S = \frac{C_4}{C_5} = -i\times
	\frac{4e^{2\eta}\cos\zeta + i\sin\zeta}
	{4e^{2\eta}\cos\zeta - i\sin\zeta},
\end{equation}
where
%
\begin{eqnarray}
 \zeta & = & \int_{x_0}^{x_1}\sqrt{|T|}dx -\frac{\pi}{4},\nonumber\\
 \eta & = & \int_{x_1}^{x_2}\sqrt{|T|}dx.
\end{eqnarray}
In the limit $e^{\eta}\to\infty$, $S$ approaches to $-i$ unless 
$\cos\zeta =0$, in which case $S=+i$. Hence, a resonance will occur
near a frequency corresponding to $\cos\zeta =0$. We denote the
value of $y$ at which 
%
\begin{equation}
 \zeta = \left( n+\frac{1}{2}\right)\pi	
	\label{eqn:quasi-bound-state}
\end{equation}
by $y_n$. Thence, we expand $S$ near $y=y_n$:
%
\begin{equation}
 S = -i\times\frac{y-y_n+ie^{-2\eta}/4\alpha_n}
	{y-y_n-ie^{-2\eta}/4\alpha_n} + O(y-y_n)^2,
	\label{eqn:S-matrix1}
\end{equation}
where
%
\begin{equation}
 \alpha_n = -\frac{d}{dy}\left.\left[
	\int_{x_0}^{x_1}\sqrt{|T|}dx\right]\right|_{y=y_n}.
\end{equation}
Because of the behavior (\ref{eqn:limit-V}), the asymptotic behavior
of $\alpha_n$ in the limit of $x_0\to -\infty$ can be
obtained~\footnote{
Although $\lambda$ in $V_{\pm}$ depends on $y$ through the eigen
equation (\ref{eqn:eq-S}), this asymptotic behavior of $\alpha_n$ is
correct since the right hand side of (\ref{eqn:limit-V}) is
independent of $\lambda$.
}:
%
\begin{equation}
 \alpha_n \sim -x_0 \sim
	\frac{\mu Mr_+}{\sqrt{M^2-a^2}}
	\left|\ln\left(\frac{r_0-r_+}{r_+-r_-}\right)\right|
	= \frac{\mu}{2\kappa}
	\left|\ln\left(\frac{r_0-r_+}{r_+-r_-}\right)\right|,
	\label{eqn:asymptotic-alpha}
\end{equation}
where $\kappa$ is 'the surface gravity of the horizon'~\footnote{
Strictly speaking, in our background there is no horizon by
assumption. However, redshifted local acceleration is bounded from
above by the surface gravity of the horizon in the extended spacetime
which has a horizon.}. 
This implies that $\alpha_n>0$. Hence, from Eq.~(\ref{eqn:S-matrix1})
and the last of Eq.~(\ref{eqn:WKBsol1}), we can conclude that there
are two regular solutions, whose asymptotic forms in the limit
$x\to\infty$ are 
%
\begin{equation}
 u \sim \exp\left[\left(\pm iy_n- \frac{1}{4\alpha_n}e^{-2\eta}
	\right)x\right].
	\label{eqn:asymptotic-u}
\end{equation}
Here the plus sign corresponds to $S^{-1}=0$ and the minus sign
corresponds to $S=0$. 
Thus, we have obtained a set of complex-frequency modes corresponding
to 
%
\begin{equation}
 \omega = \mu \left( y_n \pm\frac{i}{4\alpha_n}e^{-2\eta}\right).
\end{equation}
However, from the asymptotic behavior (\ref{eqn:asymptotic-alpha}) we
can conclude that 
%
\begin{equation}
 T_{BH}^{-1}\Im\omega \sim 
	\pm \pi e^{-2\eta}
	\left|\ln\left(\frac{r_0-r_+}{r_+-r_-}\right)\right|^{-1}
	\to 0 \quad ( r_0\to r_+ ),
\end{equation}
where $T_{BH}=\kappa /2\pi$ is the Hawking temperature of the Kerr
background.

(ii) $y>V_+(x_0)$\\
In this regime, analysis depend on how many solutions $y=V_+$ has. 
If $y=V_+$ has no solution or only one degenerate solution then there
is no complex frequency modes since whole region, $x_0\leq x$, is
classically allowed region. 
If $y=V_+$ has two solutions then we can repeat the above procedure
for the regime (i). However, obtained WKB solutions have the
asymptotic form (\ref{eqn:asymptotic-u}) with negative
$\alpha_n$ in this case. Thus, there is no regular solutions which
correspond to complex-frequency modes.

(iii) $y<-1$\\
Also in this regime, we can repeat the above procedure. If $y=V_-$ has
no solution or only one degenerate solution then there is no complex
frequency mode. 
For the case in which $y=V_-$ has two solutions, obtained WKB solution
$u$ has the asymptotic form 
%
\begin{equation}
 u \sim \exp\left[\left(\pm iy_n + \frac{1}{4\alpha_n}e^{-2\eta}
	\right)x\right].
\end{equation}
with positive $\alpha_n$. Thus, there is no regular solution which
corresponds to complex-frequency modes.

(iv) $V_-(x_0)\leq y\leq V_+(x_0)$\\
In this regime, let the solution of $y=V_+$ be $x=x_2(y)$. In each
region separated by $x_2$, the WKB solution is given by 
%
\begin{eqnarray}
 u & = & \frac{C_1}{|T|^{1/4}}
	\sinh\left(\int_{x_0}^x\sqrt{|T|}dx\right)
	\quad\mbox{for}\ x_0<x<x_2,	\nonumber\\
 u & = & \frac{C_4}{|T|^{1/4}}
	\exp\left(i\int_{x_2}^x\sqrt{|T|}dx\right)
	+\frac{C_5}{|T|^{1/4}}
	\exp\left(-i\int_{x_2}^x\sqrt{|T|}dx\right)
	\quad\mbox{for}\ x_2<x,
\end{eqnarray}
Hence, the standard connection formula for WKB solutions gives
%
\begin{equation}
 S = \frac{C_4}{C_5} = -i\times\frac{2e^{2\eta}-i}{2e^{2\eta}+i}, 
\end{equation}
where
%
\begin{equation}
 \eta = \int_{x_0}^{x_2}\sqrt{|T|}dx.
\end{equation}
From this expression of $S$, it is evident that there is no
complex-frequency mode near the real axis.

(v) $-1\leq y\leq 1$\\
In this regime, the region with large $x$ is classically forbidden
region. Thus, there is no complex-frequency mode.

Next, let us consider the case in which $ma<-2\mu Mr_+$. The above
analysis can be applied to this case by simply replacing $y$ with
$-y$, $V_+$ with $-V_-$, and $V_-$ with $-V_+$. Complex frequency
modes arise only in the regime $V_+(x_0)<y<-1$ and 
%
\begin{equation}
 T_{BH}^{-1}\Im\omega \sim 
	\pm \pi e^{-2\eta}
	\left|\ln\left(\frac{r_0-r_+}{r_+-r_-}\right)\right|^{-1}
	\to 0 \quad ( r_0\to r_+ ).
\end{equation}

Finally, let us consider the case in which 
$-2\mu Mr_+\leq ma\leq 2\mu Mr_+$. From the above analysis for other
cases, it is evident that there arise no complex-frequency mode
functions since neither the regime $1<y<V_-(x_0)$ nor the regime
$V_+(x_0)<y<-1$ exist in this case.

In summary, in this section we have shown that in the Kerr background
without horizon but with an inner boundary a scalar field has
complex-frequency modes and that the imaginary part of the complex
frequency is small enough compared with the Hawking temperature if the
inner boundary is sufficiently close to the horizon, say at a proper
altitude of Planck scale.


\section{Summary and discussion}
	\label{sec:summary}

We had analyzed the stability of the brick wall model in a rotating
background. We had shown that in the Kerr background without 
horizon but with an inner boundary a scalar field has
complex-frequency modes and that, however, the imaginary part of the
complex frequency can be small enough compared with the Hawking
temperature if the inner boundary is sufficiently close to the
horizon, say at a proper altitude of Planck scale. Hence, the time
scale of the ergoregion instability is much longer than the relaxation
time scale of the thermal state with the Hawking temperature. In the
latter time scale ambient fields should settle in the thermal
state. In this sense ergoregion instability is not so
catastrophic. Thus, the brick wall model is well defined if the inner
boundary is sufficiently close to the horizon.

Now, let us discuss physical interpretation of the existence of
complex-frequency modes. 
First, for a rotating black hole background, there is no complex
frequency mode~\cite{Whiting}. However, it is well known that
superradiant modes of fields are amplified by scattering. 
On the other hand, for the brick wall (or a rapidly rotating star)
background, the amplification of superradiant modes can be suppressed
by a boundary condition say, the Dirichlet boundary condition at the
inner boundary~\cite{MDO}.  
In stead of the amplification of superradiant modes, as shown in this 
paper, for this background there appear complex frequency modes. These
complex frequency modes are outgoing and incoming normal modes of the
field and may be understood intuitively as ``quasi-bound states'' in
the ergoregion. (See Eq.~(\ref{eqn:quasi-bound-state}).) This
interpretation is consistent with the fact that for a black hole
background there is no complex frequency mode since there is no
"quasi-bound state" in the ergoregion: any excitations with negative
energy w.r.t. observers at infinity will fall into the hole. Based on
this observation, thus, it is expected that in the brick wall
background the imaginary part of the complex frequency should become
arbitrarily small in the limit that the inner boundary becomes close
enough to the horizon since in this limit there appears a large room
for the excitations with negative energy w.r.t. observers at infinity
to escape to. This consideration is, off course, consistent with our
result in this paper: we have shown that the imaginary part of the
complex frequency can be small enough compared with the time scale
determined by the Hawking temperature if the inner boundary is
sufficiently close to the horizon, say at a proper altitude of Planck
scale.

Next, let us discuss a relation to the so called
Schiff-Snyder-Weinberg effect~\cite{SSW}. In Ref.~\cite{Fulling} the
relation between the Klein paradox~\cite{Klein} and superradiance in
a rotating black hole background was discussed in detail by using a
rectilinear model of the Kerr spacetime. Since the situation in the
Klein paradox can be understood as a limit (the so called Klein limit)
of the Schiff-Snyder-Weinberg effect~\cite{Fulling}, the situation
in a rotating black hole background should be understood as a limit of
our situation, i.e. the brick wall model. This is actually the case:
in the $x_0\to -\infty$ limit the complex frequencies in the brick
wall background disappear as in the Kerr black hole or in the
rectilinear spacetime considered in Ref.~\cite{Fulling}.

Finally, we would like to mention a possibility to stabilize the
quantum field theory in a rotating background by introducing a
nonlinear interaction. In Refs.~\cite{Migdal,Klein&Rafelski} it was
suggested that a nonlinear interaction will prevent the vacuum from
being unstable even if there are complex-frequency modes. 
It will be interesting to investigate such a possibility in the case
of quantum field theory in a rotating background. Physics in a rapidly
rotating background spacetime will be as interesting as physics of
strong fields~\cite{Greiner}.

\begin{acknowledgments}

The author would like to thank Professor W. Israel for his warmest
hospitality in University of Victoria and many truly stimulating
discussions. He also would like to thank Professor H. Kodama for his
continuing encouragement. This work was supported partially by the
Grant-in-Aid for Scientific Research Fund (No. 9809228). 

\end{acknowledgments}


\appendix


\section{Klein-Gordon norm and integrability}
	\label{app:KGnorm}

In Sec.~\ref{sec:rotating-background} we have expanded the field
operator $\phi$ as (\ref{eqn:phi-decomp}) by mode functions
$\{\Phi_{\omega lm}\}$ satisfying
(\ref{eqn:Phi-Phi=delta}-\ref{eqn:Phi*-Phi*=-delta}). 
In other words, we have required that coefficients of annihilation
operators should have positive Klein-Gordon norm instead of requiring
positivity of the frequency $\omega$ with respect to the Killing time
$t$.

In this appendix we show that the positivity of the Klein-Gordon norm
is required by integrability of equations for the ground state, say,
Eq.~(\ref{eqn:EOG}).

Let us consider a scalar field $\phi$ described by the action 
(\ref{eqn:action}) in a general $n$-dimensional globally-hyperbolic
spacetime:
%
\begin{equation}
 ds^2 = -N^2dt^2 + \gamma_{jk}(dx^j+\beta^jdt)(dx^k+\beta^kdt),
\end{equation}
where the lapse function $N$, the shift vector $\beta^j$, the
$(n-1)$-dimensional metric $\gamma_{jk}$ and the mass $\mu$ can
depend on both $t$ and the $(n-1)$-dimensional coordinates $\{x^j\}$.

To make our arguments definite, let us discretize the system of the
scalar field. Since the Lagrangian $L$ defined by $I=\int dtL$ can be
written as 
%
\begin{equation}
 L = \frac{1}{2}\int d^{n-1}x\left[\frac{\sqrt{\gamma}}{N}
	(\partial_t\phi-\beta^j\partial_j\phi)^2 
	- N\sqrt{\gamma}(\gamma^{jk}\partial_j\phi\partial_k\phi 
	+ \mu^2\phi^2)\right],
\end{equation}
we can discretize it to obtain~\footnote{
Although explicit forms of the matrices $G$, $f$ and $V$ depend on the 
way of discretization, all we need to take the continuous limit in the 
following arguments is the correspondence (\ref{eqn:correspondence})
only. Thus, the result in this appendix is independent of an explicit
way of discretization.}
%
\begin{equation}
 L = \frac{1}{2}G_{AB}(\dot{\phi}^A-f^A_C\phi^C)
	(\dot{\phi}^B-f^B_D\phi^D) - \frac{1}{2}V_{AB}\phi^A\phi^B, 
\end{equation}
provided that we suppose the following correspondence. 
%
\begin{eqnarray}
 \int d^{n-1}x\frac{\sqrt{\gamma}}{N}(\partial_t\phi)^2 
	& \Leftrightarrow & 
	G_{AB}\dot{\phi}^A\dot{\phi}^B,\nonumber\\
 \beta^j\partial_j\phi
	& \Leftrightarrow & 
	f^A_C\phi^C,\nonumber\\
 \int d^{n-1}xN\sqrt{\gamma}(\gamma^{jk}\partial_j\phi\partial_k\phi 
	+ \mu^2\phi^2)
	& \Leftrightarrow & 
	V_{AB}\phi^A\phi^B.	\label{eqn:correspondence}
\end{eqnarray}
These relations will be used when we take a continuous limit. 
The corresponding equation of motion and the commutation relations are 
%
\begin{eqnarray}
 (G_{AB}\dot{\phi}^B-G_{AB}f^B_C\phi^C\dot{)}
	+ G_{BC}f^C_A\dot{\phi}^B 
	+ (V_{AB}-f^C_AG_{CD}f^D_B)\phi^B = 0, 
	\label{eqn:EOM}
\end{eqnarray}
and 
%
\begin{equation}
 [\phi^A,\pi_B]=i\delta^A_B,\quad
 [\phi^A,\phi_B]=[\pi^A,\pi_B]=0,
\end{equation}
where the momentum $\pi_A$ conjugate to $\phi^A$ is defined by 
%
\begin{equation}
 \pi_A \equiv \frac{\partial L}{\partial\dot{\phi}^A}
	= G_{AB}(\dot{\phi}^B-f^B_C\phi^C).
\end{equation}

Now let us expand the field operator by ``mode-functions''
$\{\Phi_n^A,{\Phi_n^A}^*\}$ ($n=1,2,\cdots$), i.e. solutions of the
equation of motion: 
%
\begin{equation}
 \phi^A = \sum_n(a_n\Phi_n^A + a_n^{\dagger}{\Phi_n^A}^*), 
\end{equation}
where we assume that $\{\Phi_n^A,{\Phi_n^A}^*\}$ forms a complete set
of linearly independent solutions of the equation of motion.

We would like to define the corresponding ground state by 
%
\begin{equation}
 a_n |0\rangle =0,\quad\mbox{for}\quad\forall n.
	\label{eqn:def-0}
\end{equation}
This is the discretized version of Eq.~(\ref{eqn:EOG}). 
However, this equation is not integrable in general. Hence, in this
appendix we would like to seek the necessary and sufficient condition
for the integrability of this equation.

First, it is easily shown that a certain linear combination of
$\phi^A$ and $\pi_A$ is written as a linear combination of $a_n$ as
follows. 
%
\begin{equation}
 \pi_A -i\Omega_{AB}\phi^B =
	-i\sum_na_n\Phi^B_n(\Omega+\Omega^*)_{AB}, 
	\label{eqn:pi-phi-a}
\end{equation}
where the matrix $\Omega_{AB}$ is defined by
%
\begin{equation}
 \Omega_{AB} = -iG_{AC}[
	(\Phi^{-1})_B^{n*} \dot{\Phi}_n^{C*} -f_B^C]. 
\end{equation}
Thus, in order for Eq.~(\ref{eqn:def-0}) to be integrable, it is
necessary that 
%
\begin{equation}
 [(\pi_A -i\Omega_{AC}\phi^C),(\pi_B -i\Omega_{BD}\phi^D)]|0\rangle
	=  (\Omega_{AB}-\Omega_{BA})|0\rangle = 0,
\end{equation}
which is equivalent to 
%
\begin{equation}
 {\Phi_n^A}^*(\Omega_{AB}-\Omega_{BA}){\Phi_m^B}^* 
	= -iG_{AB}[{\Phi_n^A}^*
	(\dot{\Phi}_m^{B*}-f^B_C \Phi_m^{C*})
	-(\dot{\Phi}_n^{A*}-f^A_C \Phi_n^{C*})\Phi_m^{B*}]
	= 0.	\label{eqn:local-integrability}
\end{equation}

Next, by using the relation (\ref{eqn:pi-phi-a}), the solution of
Eq.~(\ref{eqn:def-0}) is obtained at least locally as follows,
provided that the local integrability condition
(\ref{eqn:local-integrability}) is satisfied. 
%
\begin{equation}
 \langle\{\phi^A\}|0\rangle = {\cal N}\exp
	\left[-\frac{1}{2}\Omega_{AB}\phi^A\phi^B\right].
\end{equation}
In order for this wave function to be well-defined, i.e. normalizable, 
it is necessary and sufficient that the hermite matrix
$(\Omega+\Omega^{\dagger})_{AB}$ be positive definite. This condition
is restated that the matrix $X_{nm}$ defined as follows should be
positive definite.
%
\begin{equation}
 X_{nm} \equiv 
 \Phi_n^A(\Omega+\Omega^{\dagger})_{AB}\Phi_m^{B*}
	= -iG_{AB}[\Phi_n^A
	(\dot{\Phi}_m^{B*}-f^B_C \Phi_m^{C*})
	-(\dot{\Phi}_n^A-f^A_C \Phi_n^C)\Phi_m^{B*}].
	\label{eqn:def-X}
\end{equation}

In summary, in order for Eq.~(\ref{eqn:def-0}) to be integrable, it is
necessary and sufficient that the condition
(\ref{eqn:local-integrability}) is satisfied and that the matrix 
$X_{nm}$ defined by (\ref{eqn:def-X}) is positive definite. These two
conditions can be restated as follows:
%
\begin{eqnarray}
 (\Phi_n^*,\Phi_m) & = & 0, 
	\quad\mbox{for}\quad\forall n,m,\label{eqn:integrability1}\\
 (\Phi_n,\Phi_m) & &\mbox{is positive definite},
	\label{eqn:integrability2}
\end{eqnarray}
where the norm $(\Phi,\Psi)$ is defined by
%
\begin{equation}
 (\Phi,\Psi) \equiv 	
	-iG_{AB}[\Phi^A(\dot{\Psi}^{B*}-f^B_C \Psi^{C*})
	-(\dot{\Phi}^A-f^A_C \Phi^C)\Psi^{B*}].
	\label{eqn:def-norm}
\end{equation}
It is easy to show by using the equation of motion (\ref{eqn:EOM})
that this norm is constant in time if both $\Phi$ and $\Psi$ satisfy
the equation of motion.

Provided that the condition (\ref{eqn:integrability2}) is satisfied,
it is possible to take linear transformation of $\{\Phi_n\}$ so that 
%
\begin{equation}
 (\Phi_n,\Phi_m) = \delta_{nm}, 
	\label{eqn:Phi-Phi}
\end{equation}
preserving the condition (\ref{eqn:integrability1}).
In this normalization, it can be easily shown that 
%
\begin{equation}
 [a_n,{a_m}^{\dagger}]=\delta_{nm},\quad
 [a_n,a_m]=[{a_n}^{\dagger},{a_m}^{\dagger}]=0.
\end{equation}
Thus, we can construct the Hilbert space of all quantum states spanned
by 
%
\begin{equation}
 |\{N_n\}\rangle = \left(\prod_n
	\frac{(a_n^{\dagger})^{N_n}}{N_n!}\right)|0\rangle. 
\end{equation}

Now let us take a continuous limit. 
From the correspondence (\ref{eqn:correspondence}), it is evident that 
the norm (\ref{eqn:def-norm}) is a discretized version of the
Klein-Gordon norm 
%
\begin{eqnarray}
 (\Phi,\Psi)_{KG} & = & -i\int d^{n-1}x\sqrt{\gamma}u^{\mu}
	(\Phi\partial_{\mu}\Psi^*- \Psi^*\partial_{\mu}\Phi),\\
 u^{\mu}\partial_{\mu} & = & 
	\frac{1}{N}(\partial_t -\beta^j\partial_j),\nonumber
\end{eqnarray}
which reduces to (\ref{eqn:KG-norm}) for a spacetime metric of the
form (\ref{eqn:metric}).

It is also evident that the integrability conditions
(\ref{eqn:integrability1}) and (\ref{eqn:Phi-Phi}) in the continuous
limit for the expansion (\ref{eqn:phi-decomp}) is that
(\ref{eqn:Phi-Phi*=0}) and (\ref{eqn:Phi-Phi=delta}) for $\forall
(\omega lm)\in P$ and $\forall (\omega' l'm')\in P$.


\section{Hamiltonian for a complex frequency mode}
	\label{app:complex-frequency-mode}

In this appendix, we show that it is always impossible to eliminate
terms including $a_{\lambda}^{\dagger}a_{\bar{\lambda}}^{\dagger}$
or $a_{\lambda}a_{\bar{\lambda}}$ in (\ref{eqn:H[D]}) by a Bogoliubov
transformation.

In general, a Bogoliubov transformation can be written in terms of two
matrices $\alpha$ and $\beta$ satisfying 
%
\begin{eqnarray}
 \alpha\alpha^{\dagger}-\beta\beta^{\dagger} & = & {\bf 1},\nonumber\\ 
 \alpha\beta^{T} - \beta\alpha^{T} & = & 0
	\label{eqn:cond-alpha-beta}
\end{eqnarray}
as
%
\begin{equation}
 \Phi_n \to \sum_m(\alpha_{nm}\Phi_m+\beta_{nm}\Phi^*_m),
\end{equation}
where $\{\Phi_n\}$ ($n=1,2,\cdots$) is a set of positive frequency
mode functions. Let us consider a Hamiltonian of the form
%
\begin{equation}
 H = \frac{1}{2}E^{nm}(a_na_m^{\dagger} + a_n^{\dagger}a_m)
	+ \frac{1}{2}\Lambda^{nm}a_na_m
	+ \frac{1}{2}\Lambda^{nm*}a_n^{\dagger}a_m^{\dagger},
	\label{eqn:H-general}
\end{equation}
where $E$ is a hermite matrix and $\Lambda$ is a symmetric matrix. 
Under the Bogoliubov transformation, the coefficient-matrices $E$ and
$\Lambda$ are transformed as 
%
\begin{eqnarray}
 E & \to & \left(\alpha E\alpha^{\dagger} 
	+ \beta E^*\beta^{\dagger}\right) 
	+ \left(\alpha \Lambda\beta^{\dagger} 
	- \beta \Lambda^*\alpha^{\dagger}\right),\nonumber\\
 \Lambda & \to & \left(\alpha \Lambda\alpha^T 
	- \beta \Lambda^*\beta^T\right) 
	+ \left(\alpha E\beta^T
	+ \beta E^*\alpha^T\right). 
\end{eqnarray}

Returning to the problem, the contribution of a pair of
complex-frequency modes $\lambda$ and $\bar{\lambda}$ to the
Hamiltonian (\ref{eqn:H[D]}) can be written as  
%
\begin{equation}
 h = \frac{1}{2}(\Re\omega -\Omega m)
	\left[(a_{\lambda}a_{\lambda}^{\dagger}+
	a_{\lambda}^{\dagger}a_{\lambda})
	- (a_{\bar{\lambda}}a_{\bar{\lambda}}^{\dagger}+
	a_{\bar{\lambda}}^{\dagger}a_{\bar{\lambda}})\right]
	+i\Im\omega(a_{\lambda}^{\dagger}a_{\bar{\lambda}}^{\dagger}
	-a_{\lambda}a_{\bar{\lambda}}).
	\label{eqn:sub-hamiltonian}
\end{equation}
This is of the form (\ref{eqn:H-general}) with 
%
\begin{equation}
 E = (\Re\omega -\Omega m)\left(\begin{array}{cc}
	1 & 0 \\ 0 & -1
	\end{array}\right),\quad
 \Lambda = -i\Im\omega\left(\begin{array}{cc}
	0 & 1 \\ 1 & 0
	\end{array}\right). 
\end{equation}
Hence, what we shall show now is that there is no choice of 
$2\times 2$ matrices $\alpha$ and $\beta$ satisfying
Eq.~(\ref{eqn:cond-alpha-beta}) and 
%
\begin{equation}
\left(\alpha \Lambda\alpha^T 
	- \beta \Lambda^*\beta^T\right) 
	+ \left(\alpha E\beta^T
	+ \beta E^*\alpha^T\right) = 0.
	\label{eqn:Lambda'=0}
\end{equation}
This statement is easy to show. First, since the first of
Eq.~(\ref{eqn:cond-alpha-beta}) implies that $\alpha$ has the
inverse, the second of  Eq.~(\ref{eqn:cond-alpha-beta}) and
Eq.~(\ref{eqn:Lambda'=0}) are written as  
$\gamma=\gamma^T$ and 
%
\begin{equation}
 i\Im\omega\left\{\left(\begin{array}{cc}
	0 & 1 \\ 1 & 0
	\end{array}\right)
	+ \gamma\left(\begin{array}{cc}
	0 & 1 \\ 1 & 0
	\end{array}\right)\gamma^T\right\}
 = (\Re\omega-\Omega m)\left\{\left(\begin{array}{cc}
	1 & 0 \\ 0 & -1
	\end{array}\right)\gamma^T
	+ \gamma\left(\begin{array}{cc}
	1 & 0 \\ 0 & -1
	\end{array}\right)\right\},\nonumber\\
\end{equation}
where $\gamma=\alpha^{-1}\beta$.
These equation are easy to solve with respect to
$\gamma$. The result gives $|\det\gamma|=1$. 
However, from the first of Eq.~(\ref{eqn:cond-alpha-beta}) it is
derived that $|\det\gamma|<1$, which contradicts with the above
result. Therefore, there is no choice of $2\times 2$ matrices $\alpha$
and $\beta$ satisfying Eq.~(\ref{eqn:cond-alpha-beta}) and
(\ref{eqn:Lambda'=0}).


\end{document}